# Giga-Gauss scale quasistatic magnetic field generation with laser

Ph. Korneev*

*NRNU MEPhI, Moscow 115409, Russian Federation and University of Bordeaux, CNRS, CEA, CELIA, 33405 Talence, France*

E. D'Humières, V. Tikhonchuk

*University of Bordeaux, CNRS, CEA, CELIA, 33405 Talence, France*

A simple setup for the generation of ultra-intense quasistatic magnetic fields is proposed and analysed. Estimations and numerical Particle-In-Cell calculations show that magnetic fields of gigagauss scale may be generated with conventional powerful relativistic lasers interacting with the appropriate targets of a special geometry. The setup may be useful for a wide range of applications, from laboratory astrophysics to magnetized ICF schemes.

## INTRODUCTION

Magnetic fields of different scales have been the subject of many studies since their discovery hundreds of years ago. It was then found, that astrophysical phenomena possess magnetic fields with amplitudes within a huge range from microgausses up to teragausses and even greater, deep in the relativistic region. The ultrahigh magnetic fields generation in the laboratory is a modern trend which includes the astrophysical modelling, relativistic studies with atoms and particles, Inertial Confinement Fusion (ICF), etc. Along with the utilization of the pulsing and exploding magnetic field generators, laser-assisted generation of magnetic fields attracts great interest. Modern laser facilities are powerful instruments for the generation of intense magnetic fields (see, i.e. [1–3]), as they may concentrate a lot of electromagnetic energy in small time and space regions.

In the present letter we propose a novel mechanism for the generation of the intense magnetic field, based on the direct ponderomotive electrons acceleration by a laser pulse applied to a target of a special geometry. Electron acceleration mechanism is close to that described in [4], and is accompanied with the electron guiding along the surface, irradiated by an oblique incident intense laser pulses. The effect was experimentally confirmed [5], with the laser intensity $1..2 \times 10^{18} \mathrm{W/cm}^2$, and studied numerically in different geometries [6, 7]. It was noticed, that accelerated electrons produce time-dependent currents, which may form self-consistent structures with the corresponding magnetic field. As we show below, the surface electron guiding effect may be used in quasistatic high magnetic field generation, with the field amplitudes of the order of giga-gauss level, and characteristic times of the order of at least several picoseconds. This mechanism demands, that laser intensity posesses relativistic values, and a sufficient energy deposition for achieving high amplitudes of the produced magnetic fields. Declaring the principal possibility of the setup for a laser-assisted intense magnetic field production, with the numerical modelling we show examples of the generation and the scale of the generated fields. We organize the letter as following: first, we present the results of the Particle-In-Cell simulations, then present a brief analysis of the magnetic field structure and estimate physical parameters for the possible experimental studies. Finally, we conclude by briefly discussing possible applications.

## AN EXAMPLE FOR A MAGNETIC FIELD GENERATION IN AN 'ESCARGOT' TARGET

Electrons, accelerated forward by a relativistic laser, may generate magnetic fields in Z-pinch geometry, which lives as long as the current of the accelerated particles lives. The magnetic field strength may be high in a case of high currents, but the spacial and temporary properties of these fields narrow their subsequent utilization. In the opposite, pulsed currents in a solenoid geometry [2] produce relatively long living localized magnetic fields in $\theta$-pinch geometry. Our idea is to produce the relativistic solenoid-like current of ionized electrons, for this we make use of surface electron guiding effect [4], and the plasma mirror effect [8], with a mirror of a prescribed geometry. For the current study we choosed a helix, or 'escargot', target, which is closed to a cylindrical geometry, but has a hole, for the entrance of a laser pulse, see Fig.1(A1).

$$r(\theta) = r_0 \left(1 + \frac{\delta r}{r_0} \frac{\theta}{2\pi}\right), \quad \theta \in (0, 2\pi), \qquad (1)$$

where $\theta = 0$ corresponds to the upper direction of the vertical axis in Fig.1. We examine laser-target interaction with our specific target with 2D3V Particle-In-Cell code PICLS [9]. To prove the robustness of the mechanism, we presented several runs with different target and laser parameters.

To the sake of brevity, we detaily describe only one of the runs (run (a)). In it the laser intensity was $5 \times 10^{19} \mathrm{W/cm}^2$, the laser pulse with a wavelength of 0.93 microns had a duration of 500fs. The target itself was defined by (1), with $r_0 = 43$ microns, $\delta r = 28$ microns, and it was composed with two layers by the material with ion charge $Z = 79$, ion masses $m_i^{(1)} = 36169 m_e$ of the inner 1 micron width layer, and $m_i^{(2)} = 72338 m_e$ of the outer 2 microns width layer (sublayer). The ion density was $n_i = 10^{21} \mathrm{cm}^{-3}$. Electrons with masses $m_e$ had the density $n_e = Zn_i$. The reduced ion density in this run corresponds to a possible preplasma layer on the inner surface of the target, and the more massive sublayer was introduced to save the calculation time but at the same time decrease the expansion time. The matter was presented as fully ionized cold ions, 1 particle per a cell, and 100keV hot electrons, 79 particles per a cell. The simulation box was $2160 \times 2304$ cells, or approximately $100 \times 110$ microns. The resolution was 20 steps in a time unit and 20 in a space unit. The process of the interaction is shown in Fig.1, by electron density and magnetic field $B_z$ at several subsequent time moments.

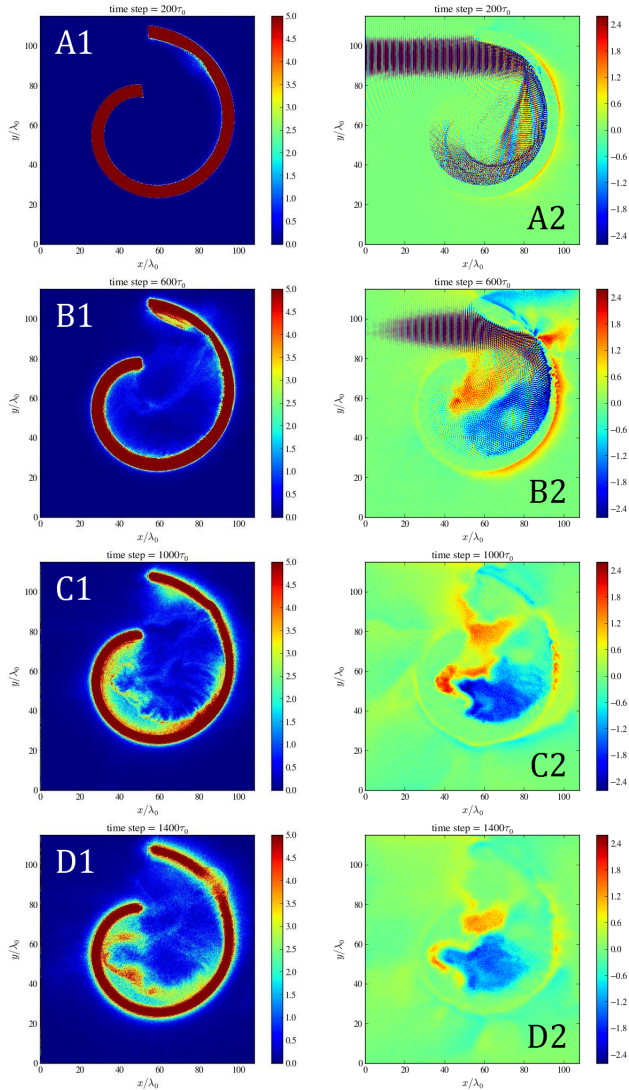

FIG. 1: Electron density (panels A1-D1), and $B_z$ field (panels A2-D2) at different time moments: 0.62, 1.9, 3.1, 4.3 ps correspondingly for A1,B1,C1,D1 and A2,B2,C2,D2, for the run (a). Electron density is shown in the units of $1.3\times10^{21} \text{cm}^{-3}$, and is cut on the value of $6.5\times10^{21}\text{cm}^{-3}$ magnetic field is shown in the units of $1.16\times10^8$ Gauss, so that maximum value of 2.6 in the colorbar corresponds to $3\times10^8$ Gauss.

As it is seen from the results of the PIC simulations in Fig. 1, even if there is a great laser energy absorption, a sufficient part of it is guided by the walls of the 'escargot' chamber, which serve as plasma mirrors of the specified form. Electrons, ablated during the laser pulse reflection from the target inner surface, are accelerated by a laser ponderomotive force along the surface, due to the surface guided mechanism, and produce a strong current, which generates a magnetic field in $z-$direction (normal to the figures plane). The generated $B_z$ field do not allow electrons to escape from the inner region of the target, forming an initial $\theta-$pinch-like structure. As time goes, TNSA mechanism [10] of ion acceleration comes into play, pulling ions from the target inner surface into the inner region. This secondary effect leads to stabilization of the $\theta-$pinch-like structure, and at the same time compresses the initial magnetic field.

To estimate a conversion efficiency from laser to the $\theta-$pinch magnetic structure energy, we consider energy balance during the interaction in Fig. 2. The total energy is increasing almost linearly with time, when laser is on, and decreases after the laser turns off. This energy loss comes mainly from the energetic electrons, which leave the simulation box. Electromagnetic energy initially also grows linearly, but approximately at 0.4 picoseconds, when the laser starts to interact with the inner surface of the target, it is absorbed by electrons. Later on, during the laser propagation, electron energy grows with time, reaching the value of the order of $\approx 60\%$ of the total energy at the end of the laser pulse (about 1.5 picoseconds). In contrary, electromagnetic energy stops to grow substantially, but after the laser is switched off it does not go to the zero value. This small part, left after 2 picoseconds, corresponds to the energy of the developed magnetic field structure. From Fig.2 it may be concluded, that after the interaction, the magnetic energy contains about $5-7\%$ of the total laser pulse energy.

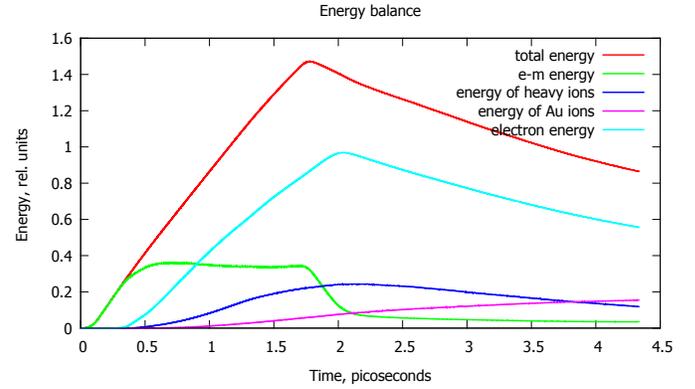

FIG. 2: Energy balance during the interaction, shown in Fig. 1 (run (a)). After 2 picoseconds, when the laser pulse is gone, electromagnetic energy is composed only by the magnetic field energy, which has the order of $5-7\%$ of the maximum (total laser pulse) energy.

We examined the robustness of the proposed mechanism by several runs with different laser and target parameters, and found the same effect. As an example we show the electron density and the magnetic field $B_z$ at a late time 6.36 picoseconds, in Fig. 3. For this run (b), the parameters we used are: the laser intensity was $10^{20}$ W/cm$^2$, the laser wavelength was 0.93 microns, the laser pulse duration was 500fs. The target with $r_0 = 100$ microns and $\delta r = 50$ microns was composed by the material with ion charge $Z = 79$, ion masses $m_i^{(1)} = 36169 m_e$, the ion density was $n_i = 2 \times 10^{22}$ cm$^{-3}$. The size of the target was approximately 2 times larger, than in the run (a). As it is seen in Fig. 3, the scale of the produced magnetic field and the magnetic structure are quite similar to that for run (a).

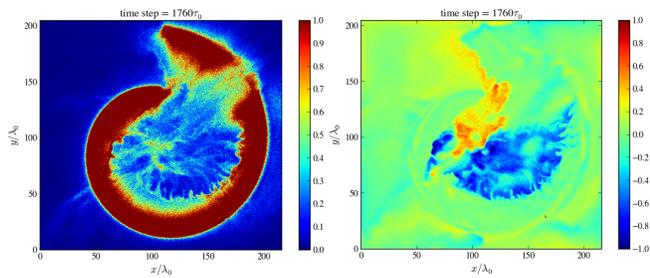

FIG. 3: Electron density (left panel), and $B_z$ field (right panel) at the time moment 6.36 ps, for the run (b). Electron density is cut on the level of $8 \times 10^{21} \text{cm}^{-3}$, magnetic field is shown in the units of $1.16 \times 10^8$ Gauss.

## DISCUSSION

The laser-target interaction and magnetic field generation may be divided in several important steps, such as electron currents generation, field formation, and magnetic structure development inside a hot cavity. The acceleration of surface electrons, lasts in run (a) from the beginning of the laser-surface interaction up to the time, when the laser pulse energy is almost totally absorbed in electron surface plasma, around $500 - 600$ fs, Fig.1(B1,B2). During this initial stage, the magnetic field strength may be roughly estimated by surface electron currents, on the base of Ampère's law. Let $\kappa$ be the conversion ratio from laser energy to energy in hot electrons, moving along the inner surface of the target, which normally is of the order of 0.1 [4]. For relativistic intensities electron velocity is close to the light velocity, then

$$B_z \sim 4\pi \kappa e n_c r^*,$$

where $e$ is the electron charge, $n_c$ is the critical electron density, $r^*$ is the characteristic length of the surface current. Considering $n_c \sim 10^{21}$ cm$^{-3}$, $r^* \sim 10$ microns, we can estimate the generated field as $B_z \sim 0.6 \times 10^9$ Gauss, the order corresponds to Fig.1(B2).

For the later stage, we mention, that depending on initial conditions, namely details of a target geometry and material, laser parameters, etc., different magnetized structures may be formed inside the cavity. Topologically in our case they may be similar to the $\theta$-pinch, which in a stationary situation can be described by the equation of the pressure balance

$$\frac{B_z^2}{8\pi} + P \approx 0, \qquad (2)$$

where $P$ is the pressure of the plasma. When the inner walls of the target are substantially heated by a laser pulse, the material of the walls ablates and an interesting effect takes place: the ablation pressure starts to compress the inner magnetic structure with hot dense plasma. This hot plasma can only slowly be mixing with the magnetic field because it rapidly becomes colisionless. However, if the absorption of the laser pulse occurs before the whole inner target surface is heated, a cold part of the surface may be magnetized. In Fig.1(C1,C2) it is seen, that to the right of the center of the target, where laser energy is high enough to produce hot plasma, magnetic field is separated from the surface plasma. In contrary, to the left of the center of the target, the surface plasma is being more magnetized during its cold stage because it can absorb enough laser energy to become collisionless later in time. So, on the second stage, the hot ablated plasma comes to the equilibrium (2) with the magnetic field inside the target hollow. During this stage, one may estimate the magnetic field from the condition, that pressure $B_z^2/8\pi$ is the same as the pressure $P = n_e T_e$. For the parameters from Fig.1 for run (a), $n_e \sim 10^{21}$ cm$^{-3}$ and $T_e \sim 1$ MeV – the ponderomotive energy for the considered laser pulse, one can get $B_z \sim 0.2 \times 10^9$ Gauss, in accordance with the value in Fig.1(D2).

The correct scaling of the presented estimations for the both stages, shows that the magnetic field amplitude is defined mainly by the laser pulse parameters. This also comes out from the comparison of different runs, as run (a) and run (b). The reason for this lies in the fact, that the ruling parameter of the irradiated solid is its electron density, which appears to be overcritical for different targets materials. The last may define, however, the life time of the generated magnetic field structure and its geometrical properties.

We considered topologically unclosed target (1), which does not allow real solenoidal currents. As we see in Fig.1 and Fig.3, in this case the generated magnetic structure has a dipole-like geometry, as it is defined by the surface electron currents. When the laser pulse inside the hollow is intense enough, though it loses a sufficient part of its energy during the reflection, it produces electron currents, which increase negative charge of the rest of the target, the left down part in our case. Return currents then appears along the target surface, on the time scale of $r_0/c$, appear where $c$ is the light velocity, and they become responsible for the magnetic field generation with the opposite direction. For the target sizes considered here, the time scale of the returning currents are of the order of 1 ps, which is also seen in Fig.1. For the generation of the monopole-type magnetic field, connected

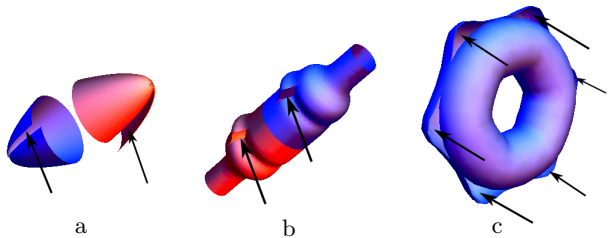

FIG. 4: Examples of target geometries for the experimental applications of the considered effect: a – two cone-like 'escargot' targets for collisions of magnetized plasmas; b – magnetic trap geometry; c – microtocamak geometry. Black arrows show laser pulses directions.

currents must be generated. This may become possible in 3D geometry, see, i.e. Fig.4 b and c.

Considering different parameters of the interaction with the 'escargot' target, we found the proposed scheme is in general robust. However, at least two competing optimizations are possible. The one is an increase of the inner target radii to increase the entrance aperture, the laser pulse inner reflection, the symmetry of the produced surface currents, and the energy conversion from the laser to magnetic field. It was shown earlier [6, 7], that surface currents may be effectively generated when laser incident angle is about 50-70 degrees. Too small target size may result in an increased laser energy absorption, without efficient directed electron currents generation. The other optimization is the decreasing of the inner target radii to increase the concentration of the magnetic field energy and to perform a stage of additional compression of the generated magnetic field by the ablated plasmas from the inner target surface. The defining parameter for the target size may thus become an entrance aperture, since it is defined by an experimental technique. In the two presented runs (a) and (b) the target size was chosen to be close to the optimum between the two considered compeeting tendencies.

## CONCLUSIONS AND PERSPECTIVES

We presented two-dimentional calculations for the high quasistationary magnetic field generation mechanism with intense laser pulses. For the possible experimental realization, it is nesessary to understand the role of 3D effects. In our simulations the intensity was of the order of $10^{19}..10^{20}$ W/cm$^2$, and the pulse duration was 0.5 ps. In terms of the laser energies, it corresponds to the order of 100 J. This is the moderate modern level scale, which makes an experimental realization feasible. However, more powerfull facilities, such as, i.e. upcoming PETAL [11], may allow the production of more intense magnetic fields.

The proposed scheme for the magnetic field generation, with the correspondent modifications, may be used in a variety of applications, such as laboratory astrophysics experiments, neutron production, different aspects of ICF, such as, i.e. electron magnetic collimation [12], etc. We discuss here several possibilities. In laboratory astrophysics applications, an intense quasistationary magnetic field generation along with the production of a highly magnetized plasma may be used, for example, for the studies of a magnetic reconnection phenomena, see, i.e. a recent review [13]. Depending on laser and target parameters it may be possible to achieve magnetized plasmas, propagating in the opposite directions, with different orientations of the magnetic field. For the standard reconnection geometry with magnetic fields in two colliding plasmas of the opposite directions, a possible example of a target is shown in Fig.4a. However, already inside a single 'escargot' target (1), with adjusted laser and target paramemeters, reconnection phenomena may be looked for, as it follows from the structure of the generated magnetic field. Another interesting application may be micrometer-scale magnetic traps, see Fig.4b and c. Target sizes and magnetic field values may be of the order of magnitude of interest for neutron production or even magnetized fusion schemes [14]. For instance, if the field values are of the order of 100MGauss, and the trap radius is about 50 microns, it can contain protons with energies $\sim 30$ MeV, and $\alpha-$particles with energies of the order of $\sim 10$ MeV. With the advanced target production technologies, it may become possible to generate toroidal magnetic structures, i.e. in target, shown in Fig.4c.

In conclusion, the idea described in the present letter is based on the ponderomotive acceleration of electrons along a target- predefined trajectory. It may work if laser intensity exceeds relativistic values. Accelerated electrons produce currents, which form a self-consistent time-dependent structure with the corresponding magnetic field. As a first "proof-of-principle" example, we show that to form a simple long-living $\theta-$pinch type electromagnetic structure, an 'escargot'-like target may be used. More complex magnetic field microstructures may be generated with certain target geometries.

## ACKNOWLEDGMENTS

Authors greatly appreciate usefull discussions with S.Fujioka and J.Santos. The work is in part supported by the French National funding agency ANR within the project SILAMPA.